%% file: paper.tex
\documentclass[sigplan,screen]{acmart}

\setcopyright{rightsretained}
\acmPrice{}
\acmDOI{10.1145/3446804.3446848}
\acmYear{2021}
\copyrightyear{2021}
\acmSubmissionID{cc21main-p39-p}
\acmISBN{978-1-4503-8325-7/21/03}
\acmConference[CC '21]{Proceedings of the 30th ACM SIGPLAN International Conference on Compiler Construction}{March 2--3, 2021}{Virtual, Republic of Korea}
\acmBooktitle{Proceedings of the 30th ACM SIGPLAN International Conference on Compiler Construction (CC '21), March 2--3, 2021, Virtual, Republic of Korea}

\usepackage{amsmath,amsfonts}
\usepackage{algorithmic}
\usepackage{graphicx}
\usepackage{textcomp}
\usepackage{xcolor}
\usepackage{listings}
\usepackage{microtype}
\usepackage[utf8]{inputenc}

\input{styles}

\begin{document}


\title{NSan: A Floating-Point Numerical Sanitizer}         


\author{Clement Courbet}
\affiliation{
  \institution{Google Research}            
  \country{France}
}
\email{courbet@google.com}          

\input{0_abstract}



\maketitle

\input{1_background}
\input{2_our_approach}

\input{3_results}

\input{4_conclusion}

\bibliography{paper}

\end{document}

%% file: styles.tex
\usepackage{listings}
\usepackage{algorithm}
\usepackage{algorithmic}
\usepackage{hyperref}

\lstdefinestyle{customcpp}{
  belowcaptionskip=1\baselineskip,
  breaklines=true,
  frame=L,
  xleftmargin=15pt,
  language=C,
  showstringspaces=false,
  basicstyle=\footnotesize\ttfamily,
  keywordstyle=\color{green!40!black},
  commentstyle=\itshape\color{purple!40!black},
  identifierstyle=\color{blue},
  numbersep=7pt,
  numbers=left,
}

\lstdefinestyle{nsanwarning}{
  belowcaptionskip=1\baselineskip,
  breaklines=true,
  frame=L,
  xleftmargin=10pt,
  showstringspaces=false,
  basicstyle=\footnotesize\ttfamily,
}

%% file: 0_abstract.tex
\begin{abstract}

Sanitizers are a relatively recent trend in software engineering. They aim at automatically finding bugs in programs, and they are now commonly available to programmers as part of compiler toolchains. For example, the LLVM project includes out-of-the-box sanitizers to detect \emph{thread safety} (\texttt{tsan}), \emph{memory} (\texttt{asan,msan,lsan}), or \emph{undefined behaviour} (\texttt{ubsan}) bugs.

In this article, we present \texttt{nsan}, a new sanitizer for locating and debugging floating-point numerical issues, implemented inside the LLVM sanitizer framework. \texttt{nsan} puts emphasis on \emph{practicality}. It aims at providing \emph{precise}, and \emph{actionable} feedback, in a \emph{timely} manner.

\texttt{nsan} uses compile-time instrumentation to augment each floating-point computation in the program with a higher-precision \emph{shadow} which is checked for consistency during program execution. This makes \texttt{nsan} between 1 and 4 orders of magnitude faster than existing approaches, which allows running it routinely as part of unit tests, or detecting issues in large production applications.

\end{abstract}

%% file: 1_background.tex
\section{Introduction}
\label{section:introduction}

Most programs use IEEE 754\cite{kahan} for numerical computation. Because speed and efficiency are of major importance, there is a constant tension between using larger types for more precision and smaller types for improved performance. Nowadays, the vast majority of architectures offer hardware support for at least 32-bit (\texttt{float})  and 64-bit (\texttt{double}) precision. Specialized architectures also support even smaller types for improved efficiency, such as \texttt{bfloat16}\cite{bfloat16}.  SIMD instructions, whose width is a predetermined byte size, can typically process twice as many \texttt{floats} as \texttt{doubles} per cycle. Therefore, performance-sensitive applications are very likely to favor lower-precision alternatives when implementing their algorithms.

Numerical analysis can be used to provide theoretical guarantees on the precision of a conforming implementation with respect to the type chosen for the implementation. However, it is time-consuming and therefore typically applied only to the critical parts of an application. To automatically detect potential numerical errors in programs, several approaches have been proposed.

\section{Related Work}
\label{section:background}

\subsection{Probabilistic Methods}

The majority of numerical verification tools use \emph{probabilistic methods} to check the accuracy of floating-point computations. They perturbate floating-point computations in the program to effectively change its output. Statistical analysis can then be applied to estimate the number of significant digits in the result. They come in two flavors: Discrete Stochastic Arithmetic (DSA)\cite{dsa} runs each floating-point operation $N$~times with a randomization of the rounding mode. Monte Carlo Arithmetic (MCA)\cite{mca} directly perturbates the input and output values of the floating-point operations.

\subsection{Manual Instrumentation}

Early approaches to numerical checking, such as CADNA~\cite{dsa}, required modifying the source code of the application and manually inserting the DSA or MCA instrumentation. While this works on very small examples, it is not doable in practice for real-life numerical applications. This has hindered the widespread adoption of these methods. To alleviate this problem, more recent approaches \emph{automatically} insert MCA or DSA instrumentation automatically.

\subsection{Automated Instrumentation}

Verificarlo~\cite{verificarlo} is an LLVM pass that intercepts floating-point instructions at the IR level (\texttt{fadd}, \texttt{fsub}, \texttt{fmul}, \texttt{fdiv}, \texttt{fcmp}) and replaces them with calls to a runtime library called \emph{backend}. The original paper describes a backend that replaces the floating-point operations by calls to an MCA library. Since the original publication, the Verificarlo has gained several backends\footnote{\url{https://github.com/verificarlo/verificarlo}}, including an improved MCA backend: \texttt{mca} is about 9 times faster than than the original \texttt{mca\_mpfr} backend\footnote{132 and 1167 ms/sample respectively for the example of section \ref{sss:kahan-sum}}.

\emph{VERROU}~\cite{verrou} and \emph{CraftHPC}~\cite{craft-hpc} are alternative tools that work directly from the original application binary. VERROU is based on the \emph{Valgrind} framework~\cite{valgrind}, while CraftHPC is based on \emph{DyninstAPI}~\cite{dyninstapi}. In both cases, the application binary is decompiled by the framework into IR, and instrumentation is performed on the resulting IR. This has the advantage that the tool does not require re-compilation of the program. However, this makes running the analysis relatively slow. In terms of instrumentation, VERROU performs the same MCA perturbation as the \texttt{mca} backend of Verificarlo, while CraftHPC detects cancellation issues (similar to Verificarlo's \texttt{cancellation} backend). A major downside of working directly from the binary is that some semantics that are available at compile time are lost in the binary. For example, the compiler knows about the semantics of math library functions such as $cos$, and knows that it has been designed for a specific rounding mode. On the other hand, dynamic tools like VERROU only see a succession of floating-point operations, and blindly apply MCA, which will result in false positives.

\subsection{Debuggability}

The main drawback of approaches based on probabilistic methods, such as Verificarlo and VERROU, is that they \emph{modify} the state of the application. Just stating that a program has numerical instabilities is not very useful, so both rely on \emph{delta-debugging}~\cite{delta-debug} for \emph{locating} instabilities. Delta debugging is a general framework for locating issues in programs based on a hypothesis-trial-result loop. Because of its generality, it is not immediately well adapted to numerical debugging. This puts a significant burden on the user who has to write a configuration for debugging\footnote{\url{https://github.com/verificarlo/verificarlo\#pinpointing-errors-with-delta-debug}}. 

FpDebug~\cite{fpdebug} takes a different approach. Like VERROU, FpDebug is a dynamic instrumentation method based on Valgrind. However, instead of using MCA for the analysis, it maintains a separate \emph{shadow value} for each floating-point value in the original application. The shadow value is the result of performing operations in higher-precision floating-point arithmetic (120 bits of precision by default). By comparing the original and shadow value, FpDebug is able to pinpoint the \emph{precise} location of the instruction that introduces the numerical error. 

%% file: 2_our_approach.tex
\section{Our Approach}

\subsection{Overview}
\label{sss-design-overview}

Based on the analysis in section \ref{section:background}, we design \texttt{nsan} around the concept of \emph{shadow values}:
\begin{itemize}
 \item Every floating-point value $v$ at any given time in the program has a corresponding \emph{shadow value}, noted $S(v)$, which is kept \emph{alongside} the original value. The shadow value $S(v)$ is typically a higher precision counterpart of $v$. A shadow value is created for every program input, and any computation on original values is applied in parallel in the shadow domain. For example, adding two values: $v_3 = add(v_1, v_2)$ will create a shadow value $S(v_3) = add_{shadow}(S(v_1), S(v_2))$, where $add_{shadow}$ is the addition in the shadow domain.
 \item At any point in the program, $v$ and $S(v)$ can be compared for consistency. When they differ significantly, we emit a warning (see section \ref{sss-precise-diagnostics}).
\end{itemize}

In our implementation, $S(v)$ is simply a floating point value with a precision that is twice that of $v$: \texttt{float} values have \texttt{double} shadow values, \texttt{double} values have \texttt{quad} (a.k.a. \texttt{fp128}) shadow values. In the special case of X86's 80-bit \texttt{long double}, we chose to use an \texttt{fp128} shadow. Note that this does \emph{not} offer any guarantees that the shadow computations will themselves be stable. However, the stability of the application computations implies that of the shadow computations, so any discrepancy between $v$ and $S(v)$ means that the application is unstable. This allows us to catch unstable cases, even though we might be missing some of them. In other words, in comparison to approaches based on MCA, we trade some coverage for speed and memory efficiency, while keeping a low rate of false positives. In our experiments, doubling the precision was enough to catch most issues while keeping the shadow value memory reasonably small.

Conceptually, our design combines the shadow computation technique of FpDebug with the compile-time instrumentation of Verificarlo. Where our approach diverges significantly from that of FpDebug is that we implement the shadow computations in LLVM IR, \emph{alongside} the original computations. This has several advantages:
\begin{itemize}
  \item \emph{Speed:} Most computations do not emit runtime library calls, the code remains local, and the runtime is extremely simple. The shadow computations are optimized by the compiler. This improves the speed by orders of magnitude (see section \ref{sss:kahan-sum}), and allows analyzing programs that are beyond the reach of FpDebug in practice (see section \ref{ssss:spec2006-perf}).
  \item \emph{Scaling: } FpDebug runs on Valgrind, which forces all threads in the application to run \emph{serially}~\footnote{\url{https://www.valgrind.org/docs/manual/manual-core.html\#manual-core.pthreads}}.
  Using compile time instrumentation means that \texttt{nsan} scales as well as the original applications. This is a major advantage in modern hardware with tens of cores.
  \item \emph{Semantics:} Contrary to dynamic approaches based on Valgrind, most of the semantics of the original program are still known at the LLVM IR stage. For example, an implementation that does not know the semantics of the program would compute the shadow of a float cosine as $S(cosf(v)) = S(cosf(S(v)))$. This would introduce numerical errors as \texttt{cosf}'s implementation is written for single-precision. Instead, \texttt{nsan} is able to replace the \texttt{cosf} by its double-precision counterpart \texttt{cos}: $S(cosf(v)) = cos(S(v))$, .
  \item \emph{Simplicity:} From the software engineering perspective, this reduces the maintenance burden by relying on the compiler for the shadow computation logic. Where FpDebug requires modified versions of the GNU Multiple Precision Arithmetic Library and GNU Multiple Precision Floating-Point Reliably in addition to the FpDebug Valgrind tool itself, in our case, LLVM handles the lowering (and potential vectorization) of the shadow code.
\end{itemize}

The following sections detail how we construct, track, and check shadow values in our implementation.

\subsection{Shadow Value Tracking}
\label{sss-value-tracking}

A \emph{floating-point value} is any LLVM value of type \texttt{float}, \texttt{double}, \texttt{x86\_fp80}\footnote{\url{https://llvm.org/docs/LangRef.html\#t-floating}}, or a~vector thereof (e.g. \texttt{<4 x float>}).

We classify floating-point values into several categories:
\begin{itemize}
 \item \emph{Temporary} values inside a function: These are typically named variables or artifacts of the programming language. They have an IR representation (and we also call them \emph{IR values}). During execution, these values typically reside within registers.
 \item \emph{Parameter} (resp. \emph{argument}) values: These are the values that are passed (resp. received) through a function call. Because numerical instabilities can span several functions, it is important that shadow values are passed to functions alongside their original counterparts.
 \item \emph{Return} values: are similar in spirit to parameter values, as the shadow must be returned alongside the original value.
 \item \emph{Memory} values are values that do not have an IR representation outside of their materialization through a \texttt{load} instruction.
\end{itemize}

\subsubsection{Temporary Values}

Temporary values are the simplest case: every IR instruction that produces a floating-point value gets a shadow IR instruction of the same opcode, but the type of the instruction is different and parameters are replaced by their shadow counterparts. We give a few examples in Table \ref{tab:nsan-def-instrumentations}.

\begin{table*}[htbp]
    \centering
    \footnotesize
    \caption{Example \texttt{nsan} instrumentation.}
    \begin{tabular}{|l|l|l|}
    \hline
    \emph{Operation} & \emph{Example} & \emph{Added Instrumentation} \\
    \hline
    binary/unary operation & \texttt{\%c = fadd float \%a, \%b} & \texttt{\%s\_c = fadd double \%s\_a, \%s\_b} \\
    \hline
    cast & \texttt{\%b = fpext <2 x float> \%a to <2 x double>} & \texttt{\%s\_b = fpext <2 x double> \%s\_a to <2 x fp128>} \\
    \hline
    select & \texttt{\%d = select i1 \%c, double \%a, double \%b} & \texttt{\%s\_d = select i1 \%c, fp128 \%s\_a, fp128 \%s\_b} \\
    \hline
    vector operation & \texttt{\%c = shufflevector <2 x float> \%a,} & \texttt{\%s\_b = shufflevector <2 x double> \%s\_a,} \\
                     & \texttt{ <2 x float> \%b, <2 x i32> <i32 1, i32 3>} & \texttt{ <2 x double> \%s\_b, <2 x i32> <i32 1, i32 3>} \\
    \hline
    known function call & \texttt{\%b = call float @fabsf(float \%a)} & \texttt{\%s\_b = call double @llvm.fabs.f64(double \%s\_a)} \\
    \hline
    fcmp & \texttt{\%r = fcmp oeq double \%a, 1.0} & \texttt{\%s\_r = fcmp oeq fp128 \%s\_a, 1.0} \\
         &                                         & \texttt{\%c = icmp eq i1 \%r, \%s\_r} \\
         &                                         & \texttt{br i1 \%c, label 2, label 1} \\
         &                                         & \texttt{1:} \\
         &                                         & \texttt{call void @\_\_fcmp\_fail\_double(...)} \\
         &                                         & \texttt{br label 2} \\
         &                                         & \texttt{2:} \\
    \hline
    return & \texttt{ret float \%a} & \texttt{store i64 i64 \%fn\_addr, i64* @\_\_ret\_tag, align 8} \\
           &                        & \texttt{\%rp = bitcast ([64 x i8]* @\_\_ret\_ptr to double*)} \\
           &                        & \texttt{store double \%s\_a, double* \%rp, align 8} \\
    \hline
    function call & \texttt{\%a = call float @returns\_float()} & \texttt{\%tag = load i64, i64* @\_\_ret\_tag, align 8} \\
                  &                                             & \texttt{\%fn\_addr = ptrtoint (float ()* @returns\_float to i64)} \\
                  &                                             & \texttt{\%m = icmp eq i64 \%tag, i64 \%fn\_addr} \\
                  &                                             & \texttt{\%rp = bitcast ([64 x i8]* @\_\_ret\_ptr to double*)} \\
                  &                                             & \texttt{\%l = load double, double* \%rp), align 8} \\
                  &                                             & \texttt{\%e = fpext float \%a to double} \\
                  &                                             & \texttt{\%s\_a = select i1 \%m, double \%l, \%e} \\
    \hline
    \end{tabular}
    \label{tab:nsan-def-instrumentations}
\end{table*}

\subsubsection{Parameter and Return Values}

Parameter values are maintained in a shadow stack. During a function call, for each floating-point parameter $v$, the caller places $S(v)$ on the shadow stack before entering the call. On entry, the callee loads $S(v)$ from the shadow stack. The only complexity comes from the fact that a non-instrumented function can call an instrumented function. Blindly reading from the shadow stack in the callee would result in garbage shadow values. To avoid this, the shadow stack is \emph{tagged} by the address of the callee. Before calling a function \texttt{f}, the caller tags the shadow stack with \texttt{f}. When reading shadow stack values, the callee checks that the shadow stack tag matches its address. If it does, the shadow values are loaded from the shadow stack. Else, the parameters are extended to create new shadows. In practice, the introduced branch does not hurt performance as it's typically perfectly predicted. 

Return values are handled in similar manner. The framework has a return slot with a tag and a buffer. Instrumented functions that return a floating-point value set the tag to their address and put the return value in the shadow return slot. Instrumented callers check whether the tag matches the callee and either read from the shadow return slot or extend the original return value (see Table~\ref{tab:nsan-def-instrumentations}). Note that because the program can be multithreaded, the shadow stack and return slot are thread-local.

\subsubsection{Memory Values}

These are a bit special because they do not have a well-defined lifetime and can persist for the lifetime of the program.

\textbf{Shadow Memory}: Like most LLVM sanitizers, we maintain a \emph{shadow memory} alongside the main application memory. The \texttt{nsan} runtime intercepts memory functions (e.g. \texttt{malloc}, \texttt{realloc}, \texttt{free}). Whenever the application allocates a memory buffer, a corresponding shadow memory buffer is allocated. The shadow buffer is released when the application buffer is released. The shadow memory is in a different address space than that of the application, which ensures that shadow memory cannot be tampered with from the application. Shadow memory is conceptually very simple: for every floating point value $v$ in application memory at address $A(v)$, we maintain its shadow $S(v)$ at address $M_s(A(v))$. A load from $A(v)$ to create a value $v$ is instrumented as a shadow load from $M_s(A(v))$ to create $S(v)$; a store to $A(v)$ creates a shadow store of $S(v)$ to $M_s(A(v))$.

\textbf{Shadow Types}: We have to handle an extra complexity: memory is untyped, so there is no guarantee that the application does not modify the value at $A(v)$ through non-floating-type stores or partial overwrites by another float. Consider the code of Fig.~\ref{fig:untyped-memory}, which modifies the byte representation of a floating-point value in memory. It's unclear how this should translate in the shadow space. In that case, we choose to resume computations by re-extending the original value: $S(*f) = *f$.

\begin{figure}
\centering
\begin{lstlisting}[style=customcpp]
float UntypedMemory(float* f) {
  *f = 1.0;
  *((char*)f + 2) = 2;
  return *f;
}
\end{lstlisting}
\caption{A function that modifies the binary representation of a floating point value in memory. How to extend this operation to the shadow domain is unclear.}
\label{fig:untyped-memory}
\end{figure}

To handle this case correctly, we track the \emph{type} of each byte in application memory. We maintain \emph{shadow types} memory. For a floating point value $v$ in application memory at address $A(v)$, each byte in the \emph{shadow types} memory at address $M_t(A(v)) + k$ contains the type of the floating point value (\texttt{unknown}, \texttt{float}, \texttt{double}, \texttt{x86\_fp80}), as well as the position $k$ of the byte within the value (see Fig.~\ref{fig:shadow-memory-layout}). A shadow value in memory is valid only if the shadow type memory contains a complete position sequence \texttt{[0,...,sizeof(type)-1]}, of the right type.  

\begin{figure}
\centering
\begin{lstlisting}[style=nsanwarning]
0x00123400:    f0 f1 f2 f3 d0 d1 d2 d3
0x00123408:    d4 d5 d6 d7 __ __ __ __
0x00123410:    l0 l1 l2 l3 l4 l5 l6 l7
0x00123418:    l8 l9 __ __ __ __ __ __
0x00123420:    d0 d1 d2 f0 f1 f2 f3 d7
0x00123428:    f0 f1 f2 f3 f0 f1 f2 f3
0x00123430:    __ __ __ __ __ __ __ __
\end{lstlisting}
\caption{Shadow type memory example: The left column is the address in application memory. For each byte in shadow type memory, the first character denotes the type: \texttt{float (f)}, \texttt{double (d)}, \texttt{long double (l)}, \texttt{unknown (\_)}; and the second character is the position of the byte inside the corresponding floating point value. In this example, the shadow memory contains valid shadows for aligned \texttt{float}s at addresses \texttt{0x00123400}, \texttt{0x00123428}, \texttt{0x0012342c}, and \texttt{0x00123423}; a \texttt{double} at address \texttt{0x00123404}; and a \texttt{long double} at address \texttt{0x00123410}. Note that the double at address \texttt{0x00123420} is not valid as it has been overwritten by the float at address \texttt{0x00123423}. }
\label{fig:shadow-memory-layout}
\end{figure}

When storing a floating point value, the shadow instrumentation retrieves the shadow pointer via a call to  a function \texttt{\_\_shadow\_ptr\_<type>\_load}, which sets the shadow memory type to \texttt{<type>} and returns the shadow value address. When loading a floating-point value, the shadow instrumentation calls a function \texttt{\_\_shadow\_ptr\_<type>\_load} which returns the shadow pointer if the shadow value is valid, and null otherwise. If the shadow is valid, it is loaded from the shadow address; else, the instrumentation creates a new shadow by extending the original load. Copying bytes from one memory location to another (either through \texttt{memcpy()} or an untyped load/store pair) copies both the shadow types and shadow values. Untyped stores and functions with the semantics of an untyped store (e.g. \texttt{memset}) set the shadow memory type to \texttt{unknown}.

In practice, subtle binary representation manipulations such as that of figure \ref{fig:untyped-memory} are very uncommon, and most untyped memory accesses fall in two categories:
\begin{itemize}
    \item Setting a memory region to a constant value (typically zero), e.g. \texttt{memset(p, 0, n * sizeof(float))}. In that case, the \texttt{nsan} framework sets the shadow types to \texttt{unknown}, and any subsequent load from this memory region will see a correct shadow value of \texttt{0}, re-extended from the original value \texttt{0}.
    \item Copying a memory region (typically, an array of floats or a struct containing a float member), e.g. \texttt{struct S \{ int32\_t i; float f; \}; void CopyS(S\& s2, const S\& s1) \{ s2 = s1; \}}. In this case, LLVM might choose to do the structure copy with a single untyped 8-byte load/store pair. \texttt{nsan} copies the shadow types from $M_t(A(s1))$ to $M_t(A(s2))$ (8 bytes) and the shadow values from $M_s(A(s1))$ to $M_s(A(s2))$ (16 bytes). Therefore, assuming that $M_t(A(s1.f))$ contains valid types, any subsequent load from \texttt{s2.f} will see the correct shadow types in $M_t(A(s2.f))$ and load the shadow value from $M_s(A(s2.f))$
\end{itemize}

In the SPECfp2006 benchmark suite, \emph{all} the floating-point loads that are done from a location with invalid or unknown types have a corresponding application value of $+0.0$, which is a strong indication that shadow types are either correctly tracked or come from an untyped store (or \texttt{memset}) of the value $0$. However, shadow type tracking is necessary for correctness and we have found it to be necessary in several places in Google's large industrial codebase.

\textbf{Memory Usage}: All allocations/deallocations are mirrored, and each original byte uses one byte in the shadow types block and two bytes in the shadow values block: quad (resp. double) is twice as big as double (resp. float). So an instrumented application uses 4 times as much memory as the original one.

\subsection{Precise Diagnostics}
\label{sss-precise-diagnostics}

We check for several types of shadow value consistency:
\begin{itemize}
\item {\emph{Observable value consistency:}} By default, we check consistency between $v$ and $S(v)$ every time a value can escape from a function, that is: function calls, return, and stores to memory. These values are the only one that are \emph{observable} by the environment (the user, or other code inside the application). This is different from the approach of FpDebug, and we'll see later that this decision has an influence on the terseness of the output and reduces false positives.
\item {\emph{Branch consistency:}} For every comparison between floating-point values, we check that the comparison of the shadow values yields the same result. This catches the case when, even though the values are very close, they can drastically affect the output of the program by taking a different execution path. This approach is also implemented in Verificarlo and VERROU.
\item {\emph{Load consistency:}} When loading a floating-point value from memory, we check that its loaded shadow is consistent. If not, this means that some uninstrumented code modified memory without \texttt{nsan} being aware. This can happen, for example, when the user used hand-written assembly code which could not be instrumented. By default, this check does not emit a warning since this is typically not an issue of the code under test. It simply resumes computation with $S(v) = v$. In practice, we found that this happened extremely rarely, and we provide a flag to disable load tracking when the user knows that it cannot happen.
\end{itemize}

In each case, we print a warning with a detailed diagnostic to help the user figure out where the issue appeared. The diagnostic includes the value and its shadow, how they differ, and a full stack trace of the execution complete with symbols source code location. An example diagnostic is given in Fig.~\ref{fig:example-warning}.

\begin{figure*}
\centering
\footnotesize
\begin{lstlisting}[style=nsanwarning]
WARNING: NumericalSanitizer: inconsistent shadow results while checking store to address 0xffda3808
double       precision  (native): dec: 0.00000000000002309503  hex: 0x1.a00b086c4888f0000000p-46
__float128   precision  (shadow): dec: 0.00000000000005877381  hex: 0x8.458cb4531bef87a00000p-47
shadow truncated to double      : dec: 0.00000000000005877381  hex: 0x1.08b1968a637df0000000p-44
Relative error: 60.70% (2^51 epsilons) (6344632558530384 ULPs == 15.8 digits == 52.5 bits)
    #0 0x7f55c33486b5 in void operations_research::glop::TriangularMatrix::TransposeLowerSolveInternal<false>(operations_research::glop::StrictITIVector<gtl::IntType<operations_research::glop::RowIndex_tag_, int>, double>*) const lp_data/sparse.cc:857:32
    #1 0x7f55c3d40104 in operations_research::glop::BasisFactorization::RightSolveForProblemColumn(gtl::IntType<operations_research::glop::ColIndex_tag_, int>, operations_research::glop::ScatteredColumn*) const glop/basis_representation.cc:452:21
    [...]
    #17 0x55a0f2c43981 in main testing/base/internal/gunit_main.cc:77:10

\end{lstlisting}
\caption{An example \texttt{nsan} warning in a real application. Note that the warning pinpoints the exact location of the issue in the original source code. The full stack trace was collapsed for clarity.}
\label{fig:example-warning}
\end{figure*}

\subsection{User Control}

\paragraph{Runtime Flags} Determining whether two floating-point values are similar is a surprisingly ill-defined problem~\cite{fp-compare}. nsan implements the \emph{epsilon} and \emph{relative epsilon} strategies from ~\cite{fp-compare}, and allows the user to customize their tolerances.
 
\paragraph{Sanitizer Interface} We provide a set of functions that can be used to interact explicitly with the sanitizer. This is useful when debugging instabilities:

\begin{itemize}
 \item \texttt{\_\_nsan\_check\_float(v)} emits a consistency check of $v$. Note that this is a normal function call: the instrumentation automatically forwards the shadow value to the runtime in the shadow stack.
 \item \texttt{\_\_nsan\_dump\_shadow\_mem(addr, size)} prints a representation of shadow memory at address \texttt{[addr,\\addr+size]}. See Fig.~\ref{fig:shadow-memory-layout} for an example.
 \item \texttt{\_\_nsan\_resume\_float(v)} Resumes the computation from the original value from that point onwards:\\ $S(v) = v$.
\end{itemize}

\paragraph{Suppressions} The framework might produce false positives. This can happen, for example, when an application performs a computation that might be unstable, but has ways to check for and correct numerical stability afterwards (see section \ref{section:results}). We provide a way to disable these warnings through \emph{suppressions}. Suppressions are specified in an external file as a function name or a source filename. If any function or filename within the stack of the warning matches a suppression, the warning is not emitted. Suppressions can optionally specify whether to resume computation from the shadow or the original value after a match.

\subsection{Interacting with External Libraries}

Most applications will at one point or other make use of code that is not instrumented. This might be because they are calling a closed-source library, because they are calling a hand-coded assembly routine, or because they are calling into the \emph{C runtime library} (e.g. \texttt{memcpy()}, or for math functions. \texttt{nsan} interacts seamlessly with these libraries thanks to the shadow tagging system described in section \ref{sss-value-tracking}.

%% file: 3_results.tex
\section{Results and Discussion}
\label{section:results}

In this section, we start by taking a common example of numerical instability and compare how Verificarlo, FpDebug and \texttt{nsan} perform in terms of diagnostics and performance. Then, we show how \texttt{nsan} compares in practice on real-life applications, using the SPECfp2006 suite. In particular, we discuss how the improved speed allows us to analyze binaries that are not approachable with existing tools, while reducing the number of false positives (and therefore the burden on the user).

\subsection{An Example: Compensated Summation}
\label{sss:kahan-sum}

Summation is probably the best known example of an algorithm which is intrinsically unstable when implemented naively. Kahan's compensated summation~\cite{kahan} works around the unstabilityof the naive summation by introducing a compensation term. Example code for both algorithms can be found on Fig.~\ref{fig:kahan-sum}.

\begin{figure}
\centering
\begin{lstlisting}[style=customcpp]
float NaiveSum(const vector<float>& values) {
  float sum = 0.0f;
  for (float v : values) {
    sum += v;
  }
  return sum;
}

float KahanSum(const vector<float>& values) {
  float sum = 0.0f;
  float c = 0.0f;
  for (float v : values) {
    float y = v - c;
    float t = sum + y;
    c = (t - sum) - y;
    sum = t;
  }
  return sum;
}
\end{lstlisting}
\caption{Naive summation and Kahan compensated summation.}
\label{fig:kahan-sum}
\end{figure}

\subsubsection{Diagnostics}
\label{ssss:kahan-diagnostics}

For each tool, we ran the two summation algorithms of Fig.~\ref{fig:kahan-sum}, on the same randomly generated vector of 10M elements. A perfect tool would warn of an instability on line $4$ in the naive case. Whether it should produce no warnings in the stable case is up for debate: On the one hand, the operations on line $13$ and $15$ result in loss of precision. On the other hand, the only thing that really matters in the end is the observable output of the function.

All three tools were able to detect the numerical issue when compiled with compiler optimizations. The tools differ quite a lot in the amount of diagnostic that they produce:
\begin{itemize}
 \item Verificarlo produces an estimate of the number of correct significant digits in both modes. The number of significant digits is lower for the naive case ($5.8$ vs $7.3$), which shows the issue. By default, no source code information is provided, though the user can optionally provide a debugging script to locate the issue~\footnote{\url{https://github.com/verificarlo/verificarlo\#pinpointing-errors-with-delta-debug}}.
 \item FpDebug evaluates the error introduced by each instructions, and sorts them by magnitude. In the naive case, FpDebug reports $1,000,008$ discrepancies, the largest of which (line $4$) has a relative error of $3.6\times10^{-5}$, which is the error introduced by the summation. In the stable case, it reports $1,000,010$ discrepancies between application and shadow value, the largest 2 being on line $15$ and $13$, with errors of about  $10^{28}$ and $10^{-3}$ respectively. This makes sense because the compensation term \texttt{c} is somehow random. The relative error for \texttt{sum} is reported to be $3.3\times10^{-8}$.
 \item \texttt{nsan} produces a single warning ($10$ lines of output) in naive mode, reporting a relative error of $3.6\times10^{-5}$ on line $6$ (\texttt{return sum}). In stable mode, it produces no output. On the one hand, \texttt{nsan} avoids producing false positives in stable mode, as the temporary variables \texttt{c}, \texttt{y}, \texttt{t}, and \texttt{sum} are only checked when producing \emph{observable} value (see section \ref{sss-precise-diagnostics}). On the other hand, the diagnostic is made on the location where the observable is produced (l.$6$) instead of the specific location where the error occurs (l.$4$). We believe that while this produces less precise diagnostics, the gain in terseness (in particular, the reduction in what we argue are false positives) benefits the user experience.
\end{itemize}

\subsubsection{Single-Threaded Performance}

To detect the issue, Verificarlo needs to run the program $N$~times, where $N$ is a large number, and run analysis on the output. In the original article, the authors use $N=1000$; it's unclear how one should pick the right value of $N$. In contrast, FpDebug and \texttt{nsan} are able to detect the issue with a single run of the program, and they can pinpoint the exact location where the issue happens.

Table~\ref{tab:nsan-speed} compares the performance of running the program without instrumentation, with Verificarlo, VERROU, FpDebug, and \texttt{nsan} respectively. Simply enabling instrumentation in Verificarlo~\footnote{\texttt{libinterflop\_ieee.so}} makes the program run about 6 times slower. This is because all instrumentation is done as function calls. Before every call, registers have to be spilled to respect the calling convention. The function call additionally prevents many optimizations because the compiler does not know what happens inside the runtime. Performing the randomization on top with the MCA backend ~\footnote{\texttt{libinterflop\_mca.so --mode=mca}} makes each sample run about 40 times slower in total. The dynamic approach of FpDebug is also quite slow as it does not benefit from compiler optimizations.

In contrast, \texttt{nsan} slows down the program by a factor of $2.3$ when shadowing \emph{float} computations as \emph{double}: shadow \emph{double} computations are done in hardware, and are as fast as the original ones, and the framework adds a small overhead. When shadowing \emph{double} computations as \emph{quad}, the slowdown is around $17$: this is because shadow computations are done in software, and are therefore much slower (some architectures supported by LLVM, such as POWER9\cite{power9}, have hardware support for quad-precision floats; \texttt{nsan} would be much faster on these). Note that all these times are given per sample. A typical debugging session in Verificarlo requires running the \texttt{mca} backend for a large number of samples (the Verificarlo authors use $1000$ samples). Therefore, analyzing even this trivial program slows it down by a factor $40000$.

\begin{table}[htbp]
    \centering
    \small
    \caption{Performance of various approaches on the Kahan Sum. The second column shows the time (in milliseconds) to run one sample of the compensated sum algorithm from Fig.~\ref{fig:kahan-sum}, with 1M elements. The third and fourth columns respectively show the slowdown compared to the original program for a single sample, and for the whole analysis (using $1000$ samples for probabilistic methods). The experiment was performed on a 6-core Xeon E5@3.50GHz with 16MB L3 cache. }
    \begin{tabular}{|l|r|r|r|}
    \hline
    \emph{Version} & \emph{ms/sample} & \emph{Slowdown} & \emph{Slowdown} \\
                   &                  & (1 sample)      &          (full) \\
    \hline
    original program          & 3.3     & 1.0x    & 1.0x   \\
    Verificarlo, ieee & 18.4    & 5.6x    & 5600x  \\
    Verificarlo, mca  & 132.3   & 40.0x   & 40000x \\
    Verrou, nearest  & 96.5    & 29.2x   & 29200x \\
    Verrou, random   & 117.0   & 35.4x   & 35400x \\
    FpDebug, precision=64   & 1573.3  & 476.6x  & 476.6x \\  
    nsan (double shadow)      & 7.7     & 2.3x    & 2.3x   \\
    nsan (quad shadow)        & 56.7    & 17.2x   & 17.2x  \\
    \hline
    \end{tabular}
    \label{tab:nsan-speed}
\end{table}

\subsubsection{Multi-Threaded Performance}

\begin{figure}
\centering
\includegraphics[width=\linewidth]{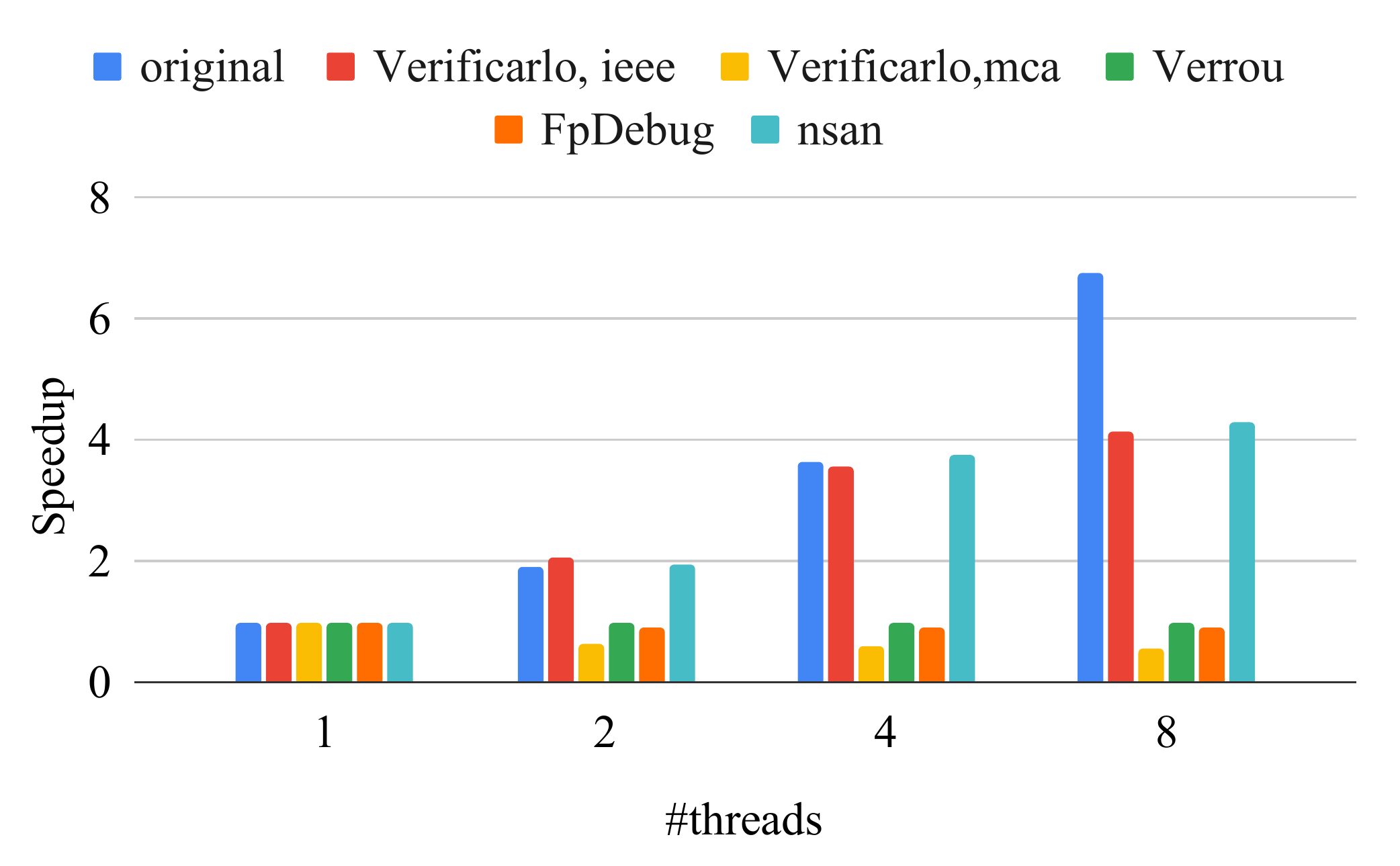}
\caption{Parallel Scalability: Speedup of running one sample of the compensated sum algorithm from Fig.~\ref{fig:kahan-sum} (100M elements) vs. number of threads.}
\label{fig:nsan-parallel-scalability}
\end{figure}

If ordering is not important, the compensated sum of Fig.~\ref{fig:kahan-sum} can be trivially parallelized: Each thread is given a portion of the array, and a last pass sums the results for each thread. Figure~\ref{fig:nsan-parallel-scalability} shows how each approach scales with the number of threads.  Because Valgrind serializes all threads, both Verrou and FpDebug cannot take advantage of additional parallelism. Methods based on compile-time instrumentation (Verificarlo and \texttt{nsan}) scale with the application. An exception is Verificarlo with the MCA backend, which is actively hurt by multithreading.

\subsection{SPECfp2006}
\label{sss:spec2006}

\subsubsection{Performance}
\label{ssss:spec2006-perf}

Table~\ref{tab:nsan-vs-fpdebug-speed} shows the time it takes to analyze each of the C/C++ benchmarks of SPECfp2006 (test set) with FpDebug and \texttt{nsan}. As shown on the simple example above, Verificarlo and Verrou take too much time to analyze large application, so we only provide compare with FpDebug. All experiments were performed on a 6-core Xeon E5@3.50GHz with 16MB L3 cache. In practice, debugging a floating-point application is likely to involve running the analysis with the application compiled in debug mode (without compiler optimizations), so we include results when the application is compiled with compiler optimizations (\emph{opt} rows) or without them (\emph{dbg} rows).

Note that all programs in SPECfp2006 are single-threaded, so this is the best case for FpDebug.

\begin{table}[htbp]
    \centering
    \small
    \caption{Performance of analyzing SPECfp2006 applications (test set) with fpdebug and nsan, with compiler optimizations turned on and off. For each experiment, we show the runtime in seconds for each and the speedup factor of nsan vs fpdebug. Note that as noted in~\cite{fpdebug}, the \texttt{dealII} benchmark cannot run under FpDebug due to limitations in Valgrind. }
    \begin{tabular}{|l|r|r|r|r|}
    \hline
    \emph{Benchmark} & \emph{Original} & \emph{FpDebug} & \emph{nsan} & \emph{Speedup} \\
    \hline
    milc (opt)    & 3.73 & 3118.2 & 505.4 &  6.2x  \\
    namd (opt)    & 8.33 & 5679.8 & 519.8 & 10.9x  \\
    dealII (opt)  & 7.60 &      - & 356.4 & -      \\
    soplex (opt)  & 0.01 &    1.9 &   0.1 & 19.0x  \\
    povray (opt)  & 0.31 &  171.8 &  12.7 & 13.5x  \\
    lbm (opt)     & 1.47 & 1343.0 & 105.4 & 12.7x  \\
    sphinx3 (opt) & 0.88 &  304.0 &  26.7 & 11.4x  \\
    \hline
    milc (dbg)     & 13.80 &  4721.1 & 502.2 &  9.4x   \\
    namd (dbg)     & 20.20 & 11445.2 & 529.0 & 21.6x   \\
    dealII (dbg)   & 85.40 &       - & 621.6 &     -   \\
    soplex (dbg)   &  0.33 &    41.0 &   0.8 & 52.5x   \\
    povray (dbg)   &  0.85 &   286.6 &  18.0 & 15.9x   \\
    lbm (dbg)      &  2.00 &  1785.0 & 105.5 & 16.9x   \\
    sphinx3 (dbg)  &  1.79 &   649.0 &  27.3 & 23.8x   \\
    \hline
    \end{tabular}
    \label{tab:nsan-vs-fpdebug-speed}
\end{table}

To investigate what made \texttt{nsan} much faster, we profiled FpDebug and \texttt{nsan} runs using the Linux perf tool~\cite{linuxperf}. Table~\ref{tab:profile} shows where the analyzed program spends most of its time. For \texttt{nsan}, we base the breakdown on calls into the compiler runtime (for \emph{quad} computation) and \texttt{nsan} runtime (shadow value load/stores and checking). This underestimates what happens in reality as the breakdown does not include additional time spent in the original application such as shadow value creation, shadow \emph{double} computations for \emph{float} values, or register spilling when calling framework functions.

For \texttt{nsan}, most time is spent on shadow computations, shadow value tracking is secondary, and checking is negligible. For FpDebug, shadow value computation (calls to \texttt{mpfr\_*}) is a much smaller part of the total. Shadow memory tracking is somehow significant, in particular the memory interceptions (calls to \texttt{vgPlain\_*}). Most time is spent executing Valgrind.

\begin{table}[htbp]
    \centering
    \small
    \caption{Approximate breakdown of where time is spent in an instrumented application (with compiler optimizations). }
    \begin{tabular}{|l|r|r|r|}
    \hline
    \emph{Benchmark} & \emph{Shadow}      & \emph{Memory}   & \emph{Value}     \\
                     & \emph{Computation} & \emph{Tracking} & \emph{Checking}  \\
    \hline
    \multicolumn{4}{|c|}{\emph{nsan}}     \\
    \hline
    milc    &  75.5\% & 4.7\%  & 0.2\%    \\
    namd    &  83.2\% & 3.0\%  &  0.7\%   \\
    dealII  &  73.7\% & 5.7\%  &  1.2\%   \\
    soplex  &  39.6\% & 11.4\% & 0.4\%    \\
    povray  &  71.8\% & 8.3\%  & 0.2\%    \\
    lbm     &  79.6\% & 2.3\%  &  1.6\%   \\
    sphinx3 &  71.2\% & 7.5\%  & 0.2\%    \\
    \hline
    \multicolumn{4}{|c|}{\emph{FpDebug}}  \\
    \hline
    milc    &  49.3\% & 14.2\% & 0.0\%   \\
    namd    &  51.6\% &  9.0\% & 0.01\%  \\
    soplex  &  14.6\% &  4.0\% & 0.1\%   \\
    povray  &  34.5\% &  7.6\% & 0.01\%  \\
    lbm     &  49.1\% & 10.6\% & 0.7\%   \\
    sphinx3 &  34.2\% &  9.7\% & 0.1\%   \\
    \hline
    \end{tabular}
    \label{tab:profile}
\end{table}

\begin{figure}
\centering
\includegraphics[width=\linewidth]{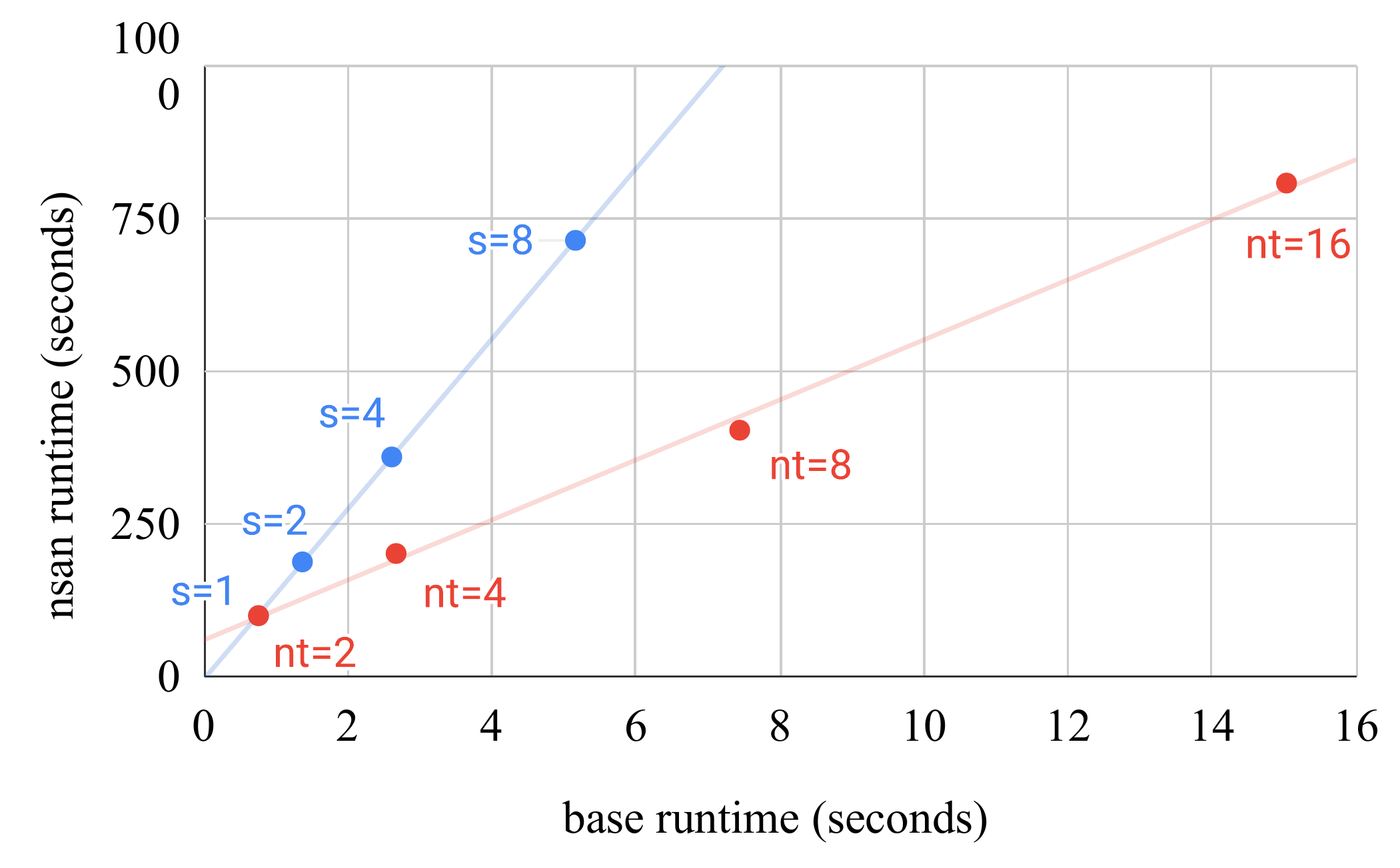}
\caption{Scaling of the \texttt{milc} benchmark with respect to problem size. We run the benchmark uninstrumented (base) and instrumented (nsan) and measure the runtime while varying the input problem size in each dimension. Each point represents a benchmark run with a particular value of \texttt{steps\_per\_trajectory} (\texttt{s}) and the grid resolution in the time domain (\texttt{nt}). Trend lines are represented for each dimension.}
\label{fig:nsan-scalability}
\end{figure}

Because \texttt{nsan} only adds a constant of work per operation, it scales linearly with respect to problem size. To assess this experimentally, we used the \texttt{milc} benchmark, which is interesting because it can scale independently in terms of memory (grid size, parameter \texttt{nt}) and number of steps (parameter \texttt{steps\_per\_trajectory}). Figure~\ref{fig:nsan-scalability} shows that \texttt{nsan} scales linearly with the problem size in both dimensions. 

\subsubsection{Diagnostics}
\label{ssss:spec2006-diagnostics}

Table~\ref{tab:positives} shows, for each tool, the number of instructions reported as introducing a relative error larger than $10^{-5}$ (a.k.a \emph{positives}). This threshold is arbitrary, and corresponds to the default for \texttt{nsan}. For this experiment, compiler optimizations are enabled as this is likely to be the configuration of choice when debugging a whole application.

\begin{table}[htbp]
    \centering
    \small
    \caption{Number of instructions introducing a relative error larger than $10^-5$. The first two columns show the number of warnings for FpDebug with and without counting the false positives from \texttt{libm}. Note that for the value marked with $^1$ the number of warnings is a lower bound as FpDebug reports unsupported vector operations \texttt{Max64Fx2} and \texttt{Min64Fx2}.}
    \begin{tabular}{|l|r|r|r|r|}
    \hline
    \emph{Benchmark}  & \emph{FpDebug} & \emph{FpDebug $\lnot$ libm} & \emph{nsan} \\
    \hline
    milc    & 140     & 0        & 0      \\
    namd    & 100     & 72       & 415    \\
    dealII  &   -     &  -       & 21     \\
    soplex  &  53     & 50       & 2      \\
    povray  & 772$^1$ & 230$^1$  & 1182   \\
    lbm     &  87     & 87       & 0      \\
    sphinx3 & 383     & 27       & 22     \\
    \hline
    \end{tabular}
    \label{tab:positives}
\end{table}

An important source of false positives for FpDebug (up to $100\%$ of the positives can be false positives) are \emph{mathematical functions} such as sine or cosine. For example, for the \texttt{milc} benchmark, \emph{all} warnings happen inside the \texttt{libm}. This is because the implementation of (e.g.) \texttt{sin(double)} uses specific constants tailored to the \texttt{double} type. Reproducing the same operations in \texttt{quad} precision is unlikely to produce a correct result. As mentioned in \ref{sss-design-overview}, LLVM is aware of the semantics of the functions of the \texttt{libc} and \texttt{libm}, which allows \texttt{nsan} to process the shadow value using the extended precision version of these functions (e.g. \texttt{sin(double)} for \texttt{sin(float)}), avoiding the false positives.

If we ignore the false positives from \texttt{libm}, \texttt{nsan} tends to reports fewer issues than FpDebug\footnote{The large number of warnings for the \texttt{namd} benchmark is due to the existence of multiple warnings inside a macro: FpDebug reports one issue for the macro, while \texttt{nsan} reports an issue for each line inside the macro. }. Unfortunately, as seen in section \ref{ssss:kahan-diagnostics}, whether a warning is a true or false positive is subject to interpretation. We inspected a sample of positives from FpDebug and \texttt{nsan}. They can roughly be classified in three buckets:
\begin{itemize}
    \item False positives due to temporary values. This is similar to the false positives in the Kahan sum from \ref{ssss:kahan-diagnostics}. These are mostly from FpDebug, though \texttt{nsan} can also produce them when memory is used as temporary value: Writing a temporary to memory makes it an \emph{observable} value. Fig.~\ref{fig:soplex-false-positive} gives examples of such a false positives.
    \item False positives due to incorrect shadow value tracking in FpDebug. FpDebug has issues dealing with integer stores that alias floating-point values in memory (a.k.a \emph{type punning}). Because \texttt{nsan} tracks shadow memory types (see \ref{sss-value-tracking})), it does not suffer from this problem. Fig.~\ref{fig:aliasing-false-positive} gives an example of this issue.
    \item Computations that are inherently unstable, and the instability is visible on a partial computation. However, the input is such that the observable output value does not differ significantly from its shadow counterpart. Fig.~\ref{fig:soplex-unclear} illustrates this. Because FpDebug checks partial computations, it warns about this case. \texttt{nsan} does not, as it only checks observables. The best tradeoff here is debatable: On one hand, the computation might become unstable with a different input. On the other hand, the code might be making assumptions about the data that the instrumentation does not know about. Until the instrumentation sees data that changes the observable behaviour of the function, it can assume that the implementation is correct.
\end{itemize}

\begin{figure}
\centering
\begin{lstlisting}[style=customcpp]
// (1).
void equal(double x, double y) {
  double d = x - y;
  if ( d > 0.00001 || d < -0.00001 ) {
    printf("error: numeric test failed! (error = %g)\n",d);
    exit(-10);
  }
}

// (2).
Real delta = 0.1 + 1.0 / thesolver->basis().iteration();
...
x = coPenalty_ptr[j] += rhoVec[j]*(beta_q*rhoVec[j]-2*rhov_1*workVec_ptr[j]);
if (x < delta)
  coPenalty_ptr[j] = delta;
\end{lstlisting}
\caption{Example false positives from the \texttt{soplex} and \texttt{namd} benchmark. For (1), note how a large relative error can be created by cancellation on line $3$. However, all that matters is the absolute value compared to $0.00001$. FpDebug incorrectly warns on that case, while \texttt{nsan} is silent. (2) is similar in spirit, though more complex. The cancellation potentially introduced on line $14$ is handled on line $15-16$, but both FpDebug and \texttt{nsan} report an issue on line $13-14$.}
\label{fig:soplex-false-positive}
\end{figure}

\begin{figure}
\centering
\begin{lstlisting}[style=customcpp]
void __attribute__((noinline)) Neg(double* v) {
  *((unsigned char*)v + 7) ^= 0x80;
}

double Example(double v) {
  double d = v / 0.2 - 3.0;
  Neg(&d);
  return d;
}

\end{lstlisting}
\caption{Example false positive with type punning. FpDebug can be made to report an arbitrarily large error, as it uses a non-negated shadow value for $S(d)$ after the call to \texttt{Neg}. In \texttt{Example}, the computation is unstable around \texttt{v=0.6}, and FpDebug returns an error of 260\% instead of the correct value of 60\%. \texttt{nsan} is able to detect that the last two bytes of the shadow value have been invalidated by the untyped store thanks to shadow type tracking. \\
Note: the code was adapted from more complex application code, \texttt{noinline} added to prevent some compiler optimizations.}
\label{fig:aliasing-false-positive}
\end{figure}

\begin{figure}
\centering
\begin{lstlisting}[style=customcpp]
// Unstable loop.
for (i = 2; i <= Octaves; i++) {
  ...
  result[Y] += o * value[Y];
  result[Z] += o * value[Z];
  if (i < Octaves) {
    ...
    o *= Omega;
  }
}

// Division by small value.
if (D[Z] > EPSILON) {
  ...
  t = (Corner1[Z] - P[Z]) / D[Z];
}
\end{lstlisting}
\caption{Example true positives from the \texttt{povray} benchmark. The loop accumulates values of widely different magnitudes, which is known to produce large numerical errors. The first one is caught only by \texttt{nsan}, likely because it's vectorized by the compiler, and FpDebug does not handle some vector constructs. Both tools catch the second. }
\label{fig:povray-true-positive}
\end{figure}

\begin{figure}
\centering
\begin{lstlisting}[style=customcpp]
Real SSVector::length2() const {
  Real x = 0.0;
  for(int i = 0; i < num; ++i)
    x += val[idx[i]] * val[idx[i]];
  return x;
}
\end{lstlisting}
\caption{Example code from the \texttt{soplex} benchmark. While the elements of the sum and the partial sum diverge from their shadow counterpart, the eventual result does not. FpDebug reports an issue on line $4$, but not on line $5$. \texttt{nsan} does not report an issue.}
\label{fig:soplex-unclear}
\end{figure}

\subsection{Limitations}

We have mentioned earlier that \texttt{nsan} only checks \emph{observable} values within a function, and we have seen previous sections that this approach helps prevent false positives. However, this also makes \texttt{nsan} susceptible to compiler optimizations such as inlining (resp. outlining). Because these optimizations change the boundaries of a function, they change its \emph{observable} values. For example, given the code of Fig.~\ref{fig:sum-inline}, a compiler might decide to inline \texttt{NaiveSum} into its caller \texttt{Print}.

\begin{figure}
\centering
\begin{lstlisting}[style=customcpp]
float NaiveSum(const vector<float>& values) {
  float sum = 0.0f;
  for (float v : values) sum += v;
  return sum;
}

void Print(const vector<float>& values) {
  float v = NaiveSum(values) + 1.0;
  printf("%f", v);
}
\end{lstlisting}
\caption{Example code where inlining might change the output of nsan. The only observable value of \texttt{NaiveSum} is its return value. The only observable value of \texttt{Print} is the second argument to the \texttt{printf} call. Depending on whether \texttt{NaiveSum} is inlined, the warning is emitted on line $6$ column $10$, or line $11$ column $16$.}
\label{fig:sum-inline}
\end{figure}

In that case, the \texttt{sum} value will not be checked by \texttt{nsan} on line $6$, because \texttt{sum} is not an observable value of \texttt{NaiveSum}. This is not an issue for detecting numerical stability, as the \texttt{sum} variable is still tracked within \texttt{Print}. However, it changes the \emph{source location} where \texttt{nsan} reports the error. While the user can easily circumvent the issue by using \texttt{\_\_nsan\_check\_float()} function to debug where the error happens exactly, this degrades the user experience as it requires manual intervention.

However, LLVM internally tracks function inlining in its debug information. In the future we plan to to correct the issue above by emitting checks for observable values of inlined functions within their callers.

%% file: 4_conclusion.tex
\section{Conclusion}

Even though \texttt{nsan} offers less guarantees than numerical analysis tools based on probabilistic methods, it was able to tackle real-life applications that are not approachable with these tools in practice due to prohibitive runtimes.

We've shown that \texttt{nsan} was able to detect a lot of numerical issues in real-life applications, while drastically reducing the number of false positives compared to FpDebug. Our sanitizer provides precise and actionable diagnostics, offering a good debugging experience to the end user.

Because \texttt{nsan} works directly in LLVM IR, shadow computations benefit from compiler optimizations, and can be lowered to native code, which reduces the analysis cost by at least an order of magnitude compared to other approaches.

We believe that user experience, and in particular execution speed and scalability, was a major factor for the adoption of toochain-based sanitizers over Valgrind-based tools, and we aim to emulate this success with \texttt{nsan}. We think that this new sanitizer is a step towards wider adoption of numerical analysis tools.

We intend to propose \texttt{nsan} for inclusion within the LLVM project, complementing the existing sanitizer suite.